\documentclass[aps,showpacs,showkeys,prd,twocolumn]{revtex4}
\usepackage{amssymb}
\usepackage{amsfonts}
\usepackage{amsmath}
\usepackage{graphicx}
\usepackage{color}
\usepackage[dvips]{epsfig}
\usepackage[dvips]{graphicx}
\usepackage{float}
 
\begin{document}

\title{Self-dual soliton solutions in a Chern-Simons-CP(1) model \\
with a nonstandard kinetic term}
\author{Rodolfo Casana and Lucas Sourrouille}
\affiliation{Departamento de F\'{\i}sica, Universidade Federal do Maranh\~{a}o,
65080-805, S\~{a}o Lu\'{\i}s, Maranh\~{a}o, Brazil.}

\begin{abstract}
A generalization of the Chern-Simons-CP(1) model is considered by introducing a nonstandard
kinetic term. For a particular case, of this nonstandard kinetic term, we show that the model
support self-dual Bogomol’nyi equations. The BPS energy has a bound proportional to the sum
of the magnetic flux and the CP(1) topological charge.
The self-dual equations are solved analytically and verified numerically.
\end{abstract}

\maketitle

\section{Introduction}

The $CP(n)$ sigma model have been investigated in detail since the early
70's mainly as toy models to explore the strong coupling effects of $QCD$
and as effective models of some condensed matter systems \cite{gral}-\cite{gral2}.

An important issue related to this type of models concern the existence of
soliton type solutions. For the simplest $CP(1)$ mode topological solutions
have been shown to exist\cite{polyakov}. Nevertheless, the solutions are of
arbitrary size due to scale invariance. As argued originally by
Dzyaloshinsky, Polyakov and Wiegmann\cite{polyakov1} a Chern-Simons term can
naturally arise in this type of models and the presence of a dimensional
parameter could play some role stabilizing the soliton solutions. A first
detailed consideration of this problem was done in Ref.\cite{voru} where a
perturbative analysis around the scale invariant solutions (i.e no Chern
Simons coupling $\kappa=0$) showed that the solutions were pushed to
infinite size. The problem of the stabilizing topological soliton of the
pure $CP(1)$ model by including a Chern-Simons and Maxwell term for the
gauge field, was done by nonperturbative analysis in Ref.\cite%
{z,ta,tay,ls,ls1,ls2}. Although, these works showed the possibility of
stabilizing the $CP(1)$ sigma model solitons, they do not present
self-duality Bogomol'nyi equations. This lack of self-duality can present
itself as a disadvantage, at least technically. One reason is that self-dual
vortices, such as, for example, those of the Abelian Higgs model \cite{SW}
do not interact by virtue of the stress tensor vanishing identically, and
hence multivortex configurations arbitrarily distributed on the plane can be
studied systematically.

In the recent years, theories with nonstandard kinetic term, named $k$-field
models, have received much attention. The $k$-field models are mainly in
connection with effective cosmological models\cite{APDM}-\cite{APDM6} as well as the
tachyon matter\cite{12}, the ghost condensates \cite{13}-\cite{13d} and dark matter\cite%
{APL}. The addition of non-linear terms to the kinetic part of the
Lagrangian has interesting consequences for topological defects, making it
possible for defects to arise without a symmetry-breaking potential term
\cite{sy}. In this context several studies have been conducted, showing that
the $k$-theories can support topological soliton solutions both in models of
matter as in gauged models\cite{BAi}-\cite{BAi4}, \cite{SG}-\cite{SG4}, \cite{LS}\cite{LS1}. These solitons have certain
features such as their characteristic size, which are not necessarily those
of the standard models\cite{B}-\cite{B2}

In this article, we investigate a Chern-Simons-$CP(1)$ model with a
nonstandard kinetic term. We will show that introducing a particular
nonstandard dynamics in a Chern-Simons-$CP(1)$ model, via a function $\omega$
depending on the $CP(1)$ field, we can obtain self-duality Bogomol'nyi
equations by minimizing the energy functional of the model. Finally,
we will be able to solve the Bogomol'nyi equations and obtain novel analytic
expressions for the soliton solutions. This analysis is completed by showing
explicitly the principal features of the soliton profiles. 

\section{The theoretical framework}

We begin by considering the following $(2+1)$-dimensional Chern-Simons
model, coupled to a complex $CP(1)$ field $n(x)$, subject to the constraint $%
n^{\dagger }n=1$
\begin{equation}
S=S_{cs}+\int d^{3}x\left( \left\vert D_{\mu }n\right\vert
^{2}-V(n,n^{\dagger })\right)  \label{S1}
\end{equation}%
where $S_{cs}$ is the Chern-Simons action,
\begin{equation}
S_{cs}=\int d^{3}x\frac{\kappa }{4}\epsilon ^{\mu \nu \rho }A_{\mu }F_{\nu
\rho }
\end{equation}%
here $F_{\mu \nu }=\partial _{\mu }A_{\nu }-\partial _{\nu }A_{\mu }$, is
electromagnetic strength-tensor, $\ D_{\mu }=\partial _{\mu }+iA_{\mu }$ is
the covariant derivative. The metric tensor is defined as $g^{\mu \nu
}=(1,-1,-1)$. The potential $V(n,n^{\dagger })$ is a function of the field $%
n(x)$ and its complex conjugate. The constraint can be introduced in the
variational process via a Lagrange multiplier. Then, we extremise the
following action
\begin{equation}
S=S_{cs}+\int d^{3}x\left( \left\vert D_{\mu }n\right\vert
^{2}-V(n,n^{\dagger })+\lambda (n^{\dagger }n-1)\right)
\end{equation}%
The variation of this action yields the field equations,
\begin{equation}
\frac{1}{2}\kappa \epsilon _{\mu \nu \rho }F^{\nu \rho }+J_{\mu }=0
\label{motion}
\end{equation}%
\begin{equation}
D_{\mu }D^{\mu }n+\frac{\partial V}{\partial n^{\dagger }}-\lambda n=0,
\end{equation}%
where $J_{\mu }=i[(D_{\mu }n)^{\dagger }n-n^{\dagger }D_{\mu }n]$ is the
conserved current density and $\lambda =n^{\dagger }\left( D_{\mu }D^{\mu }n+%
\frac{\partial V}{\partial n^{\dagger }}\right) $, so that
\begin{equation}
D_{\mu }D^{\mu }n+\frac{\partial V}{\partial n^{\dagger }}=\left[ n^{\dagger
}\left( D_{\mu }D^{\mu }n+\frac{\partial V}{\partial n^{\dagger }}\right) %
\right] n.
\end{equation}

The time component of Eq.(\ref{motion})
\begin{equation}
\kappa F_{12}=-J_{0}  \label{gauss}
\end{equation}%
is Gauss's law of Chern-Simons dynamics. Integrating it over the entire
plane one obtains the important consequence that any object with charge $Q=%
\displaystyle\int d^{2}xJ_{0}$ also carries magnetic flux $\Phi =%
\displaystyle\int Bd^{2}x$ \cite{Echarge}-\cite{Echarge2}:
\begin{equation}
\Phi =-\frac{1}{\kappa }Q,
\end{equation}%
where in the expression of magnetic flux we renamed $F_{12}$ as $B$.

The expression of the energy functional for the static field configuration
is
\begin{equation}
E=\int d^{2}x\left( \frac{\kappa ^{2}}{4}B^{2}+|D_{i}n|^{2}+V(n,n^{\dagger
})\right) .  \label{statich}
\end{equation}

As mentioned in Ref.\cite{z,ta,tay}, the model (\ref{S1}) does not support
Bogomol'nyi equations. In fact, as was shown in the Ref.\cite{z}, it can be
established that the energy functional (\ref{statich}) is bounded below by a
multiple of the winding number, which is guaranteed to be a non-vanishing
energy for non-trivial field configuration. Despite this, it is not possible
to saturate the topological bound. This is because the nature of the $CP(1)$
field, which prevents rewrite the expression (\ref{statich}) as sum of
square terms plus a topological term. We will consider, here, a
generalization of the model (\ref{S1}), which consist on a modification of
the model (\ref{S1}) by introducing a nonstandard kinetic term.
Specifically, we consider the following $(2+1)$ dimensional model with
Chern-Simons-CP(1) Lagrangian
\begin{eqnarray}
S &=&\int d^{3}x\left( \frac{\kappa }{4}\epsilon ^{\mu \nu \rho }A_{\mu
}F_{\nu \rho }+\omega (n,n^{\dagger })|D_{0}n|^{2}\right.  \notag \\
&&~\ \ \ \ \ \left. \frac{{}}{{}}-|D_{i}n|^{2}+V(n,n^{\dagger })\right)
\label{S2}
\end{eqnarray}%
Here, the function $\omega ((n,n^{\dagger }))$ is, in principle, an
arbitrary function of the $CP(1)$ field.

Since, we are interested on time-independent soliton solutions that ensure
the finiteness of the action (\ref{S2}), we are looking for stationary
points of the energy which for the static field configuration reads
\begin{equation}
E=\int \,\,d^{2}x\left( \frac{{}}{{}}-\kappa A_{0}B-\omega (n,n^{\dagger
})A_{0}^{2}+|D_{i}n|^{2}+V(n,n^{\dagger })\right)  \label{EJP}
\end{equation}%
The Gauss law (\ref{gauss}) for this system takes the form
\begin{equation}
A_{0}=-\frac{\kappa }{2}\frac{B}{\omega (n,n^{\dagger })}.  \label{A0}
\end{equation}%
Substitution of (\ref{A0}) into (\ref{EJP}) leads to
\begin{equation}
E=\int \,\,d^{2}x\left( \frac{\kappa ^{2}}{4}\frac{B^{2}}{\omega
(n,n^{\dagger })}+|D_{i}n|^{2}+V(n,n^{\dagger })\right)  \label{EJP1}
\end{equation}

In order to implement the BPS formalism we first introduce the usual ansatz
for describe $CP(1)$\ solutions%
\begin{equation}
n=\left(
\begin{array}{c}
e^{iN\phi }\cos \frac{\theta }{2} \\
\sin \frac{\theta }{2}%
\end{array}%
\right) ,~\theta =\theta \left( r\right) ,~A_{\phi }\left( r\right) =a\left(
r\right) ,
\end{equation}%
where $N$ is the winding number.

In this ansatz, the magnetic field reads as%
\begin{equation}
B=\frac{\left( ra\right) ^{\prime }}{r},
\end{equation}%
the quantity $|D_{i}n|^{2}$ is expressed as%
\begin{equation*}
\left\vert D_{i}n\right\vert ^{2}=\left( \frac{\theta ^{\prime }}{2}\right)
^{2}+\left( a+\frac{N}{r}\right) ^{2}\cos ^{2}\frac{\theta }{2}+a^{2}\sin
^{2}\frac{\theta }{2}.
\end{equation*}%
Under this prescription, the energy (\ref{EJP1}) is expressed as%
\begin{eqnarray}
E &=&\int d^{2}x\left[ \frac{\kappa ^{2}B^{2}}{4\omega \left( \theta \right)
}+V\left( \theta \right) +\left( \frac{\theta ^{\prime }}{2}\right)
^{2}\right.  \\
&&\hspace{1cm}\left. +\left( a+\frac{N}{r}\right) ^{2}\cos ^{2}\frac{\theta
}{2}+a^{2}\sin ^{2}\frac{\theta }{2}\right] .  \notag
\end{eqnarray}%
The Ampere's law reads
\begin{equation}
\frac{\kappa ^{2}}{4}\left( \frac{B}{\omega }\right) ^{\prime }-\left[ a+%
\frac{N}{2r}\left( 1+\cos \theta \right) \right] =0,  \label{eqmov1}
\end{equation}%
and the $\theta -$ field equation is

\begin{equation}
\theta ^{\prime \prime }+\frac{\theta ^{\prime }}{r}+\frac{\kappa ^{2}}{2}%
\frac{~B^{2}}{\omega ^{2}}\frac{\partial \omega }{\partial \theta }-2\frac{%
\partial V}{\partial \theta }+\left( 2a+\frac{N}{r}\right) \frac{N}{r}\sin
\theta =0.  \label{eqmov2}
\end{equation}

In the following we apply the BPS\ formalism, so after some
manipulations the energy can be expressed in the following quadratic form%
\begin{eqnarray}
E &=&\int d^{2}x\ \left\{ \frac{\kappa ^{2}}{4\omega }\left( B\pm \frac{2}{%
\kappa }\sqrt{\omega V}\right) ^{2}\mp \kappa B\sqrt{\frac{V}{\omega }}%
\right.   \notag \\
&&\hspace{1cm}+\left[ \frac{\theta ^{\prime }}{2}\sin \frac{\theta }{2}\pm
\left( a+\frac{N}{r}\right) \cos \frac{\theta }{2}\right] ^{2}\
\label{EJP1a} \\
&&\hspace{1cm}\left. +\left( \frac{\theta ^{\prime }}{2}\cos \frac{\theta }{2%
}\mp a\sin \frac{\theta }{2}\right) ^{2}\mp \frac{1}{2}\frac{N\sin \theta }{r%
}\theta ^{\prime }\ \right\} ,  \notag
\end{eqnarray}%
it is minimized by imposing
\begin{eqnarray}
B &=&\mp \frac{2}{\kappa }\sqrt{\omega V},  \label{bps1} \\
\frac{\theta ^{\prime }}{2}\sin \frac{\theta }{2} &=&\mp \left( a+\frac{N}{r}%
\right) \cos \frac{\theta }{2},  \label{bps2} \\
\frac{\theta ^{\prime }}{2}\cos \frac{\theta }{2} &=&\pm a\sin \frac{\theta
}{2},  \label{bps3}
\end{eqnarray}%
where the upper(lower) signal corresponds to $N>0$ $(N<0)$.

The two last equations can be reduced to one given
\begin{equation}
\theta ^{\prime }=\mp \frac{N}{r}\sin \theta ,  \label{BPS_1}
\end{equation}%
it will be the first BPS equation.

In order to establish a lower bound for the energy the topological
solutions, we choose the function $\omega (\theta )$ as
\begin{equation}
\omega (\theta )=\kappa ^{2}V(\theta ),  \label{omega}
\end{equation}%
so the\ BPS equation\ (\ref{bps1})\ of our CP(1) model becomes%
\begin{equation}
B=\mp 2V\left( \theta \right) .  \label{BPS_2}
\end{equation}

In addition it is interesting to note that the relations (\ref{A0}), (\ref%
{omega}) and (\ref{BPS_2}), imply
\begin{equation}
A_{0}=\pm \frac{1}{\kappa }  \label{A02}
\end{equation}%
which lead us to a soliton solution without electric charge.

In this way, the BPS energy becomes
\begin{equation}
E_{BPS}=\int d^{2}x\left( \mp B\pm \frac{N}{2}\frac{\left( \cos \theta
\right) ^{\prime }}{r}\right) ,  \label{Els}
\end{equation}%
where the first integral is the magnetic flux and the second is the CP(1)
topological charge \cite{ls}. The requirement of well-behaved fields at
origin and infinity, provide the following boundary conditions for the
fields $\theta \left( r\right) $ and $a\left( r\right) :$
\begin{equation}
\theta \left( 0\right) =\pi ~,~~\theta \left( \infty \right) =0,
\label{bctheta}
\end{equation}%
\begin{equation}
a\left( 0\right) =0~,~\ \left( ra\right) \left( \infty \right) =-C_{N}
\label{bcvector}
\end{equation}%
with $C_{N}>0\left( C_{N}<0\right) $ for $N>0~\left( N<0\right) $.

With such boundary conditions allow compute the magnetic flux
\begin{equation}
\Phi =\int \,\,d^{2}xB=-2\pi C_{N}\;,  \label{fluxx}
\end{equation}%
whereas the BPS\ energy (\ref{Els}) becomes%
\begin{equation}
E_{BPS}=2\pi \left\vert C_{N}\right\vert +2\pi \left\vert N\right\vert ,
\label{E_bound}
\end{equation}

By using BPS\ equations the energy (\ref{EJP1a}) can be written as
\begin{equation}
E_{BPS}=\int d^{2}x~\varepsilon _{BPS},
\end{equation}%
where $\varepsilon _{BPS}$ is the BPS energy density,
\begin{equation}
\ \varepsilon _{BPS}=2V(\theta )+\frac{N^{2}\sin ^{2}\theta }{2r^{2}},
\label{enbps}
\end{equation}%
it is a positive quantity.

Since the solutions of self-dual equations (\ref{BPS_1}) and (\ref%
{BPS_2}) should be also solutions of the second order equations of motion (\ref%
{eqmov1}) and (\ref{eqmov2})\ we require that the field $a(r)$ obeys%
\begin{equation}
a\left( r\right) =-\frac{N}{2r}\left[ 1+\cos \theta \left( r\right) \right]
\text{. }  \label{aaa}
\end{equation}%
This equation allows to compute the magnetic field%
\begin{equation}
B=\frac{\left( ar\right) ^{\prime }}{r}=\mp \frac{N^{2}}{2r^{2}}\sin
^{2}\theta ,
\end{equation}%
which implies an equation for the potential,
\begin{equation}
2V\left( \theta \right) = \frac{N^{2}}{2r^{2}}\sin^{2}\theta
\end{equation}
This is similar to the ``superpotential equation'' found in Ref.\cite{Ws1} and \cite{Ws2}, which relates the potential with 
topological terms.
In the following we solve the BPS equations for $N>0$. The first BPS
equation (\ref{BPS_1}),
\begin{equation}
\theta ^{\prime }=-\frac{N}{r}\sin \theta ,
\end{equation}%
is solved explicitly and its solution compatible with the boundary
conditions (\ref{bctheta}) is given by%
\begin{equation}
\theta \left( r\right) =\arctan \left( \frac{2\left( \frac{r}{r_{0}}\right)
^{N}}{\left( \frac{r}{r_{0}}\right) ^{2N}-1}\right) .
\end{equation}%
where $r_{0}$ is a parameter characterizing the effective radius of the
topological defect. Then, the equation (\ref{aaa}) reads 
\begin{equation}
a\left( r\right) =-\frac{N}{r}\frac{\left( \frac{r}{r_{0}}\right) ^{2N}}{%
\left( \frac{r}{r_{0}}\right) ^{2N}+1}
\end{equation}%
It provides the following behavior at $r=0$, %
\begin{equation}
a\left( r\right) =-\frac{N}{\left( r_{0}\right) ^{2N}}\ r^{2N-1}+...
\end{equation}%
and for $r\rightarrow \infty $, we get
\begin{equation}
a\left( r\right) =-\frac{N}{r}+N\left( r_{0}\right) ^{2N}\frac{1}{r^{2N+1}}%
+...,  \label{r-inf}
\end{equation}%
it allows to determine asymptotic constant $C_{N}$,
\begin{equation}
C_{N}=N,  \label{ccn}
\end{equation}%
This implies the magnetic flux and BPS energy density are
proportional to $N$.

We shown the profiles of the BPS solutions for $r_{0}=5$, and some values of the winding number $N$.

\begin{figure}[tbp]
\centering\includegraphics[width=8.5cm]{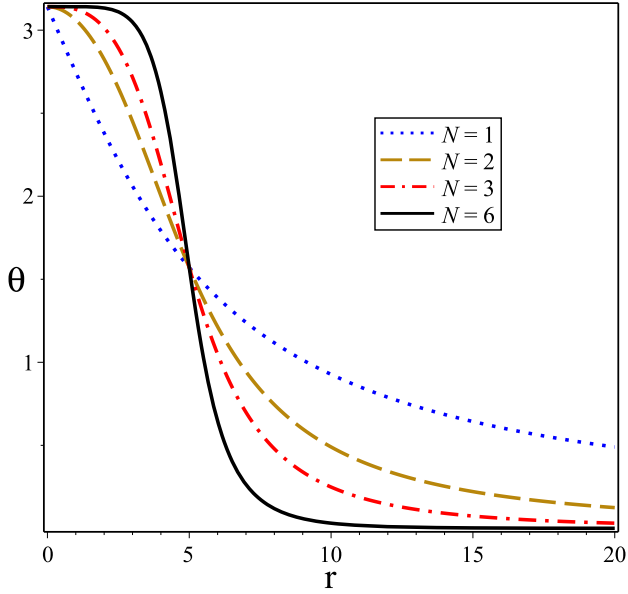}
\caption{$\protect\theta \left( r\right) $ field profiles.}
\label{gg}
\end{figure}
The Fig. \ref{gg} shows the profiles of the $\theta $ field. It is
clear that for $N>1$, the asymptotic values is attained rapidly.
For larger $N$, the profiles is a rectangle with height $\pi$ and width $r_{0}$.

\begin{figure}[tbp]
\centering\includegraphics[width=8.5cm]{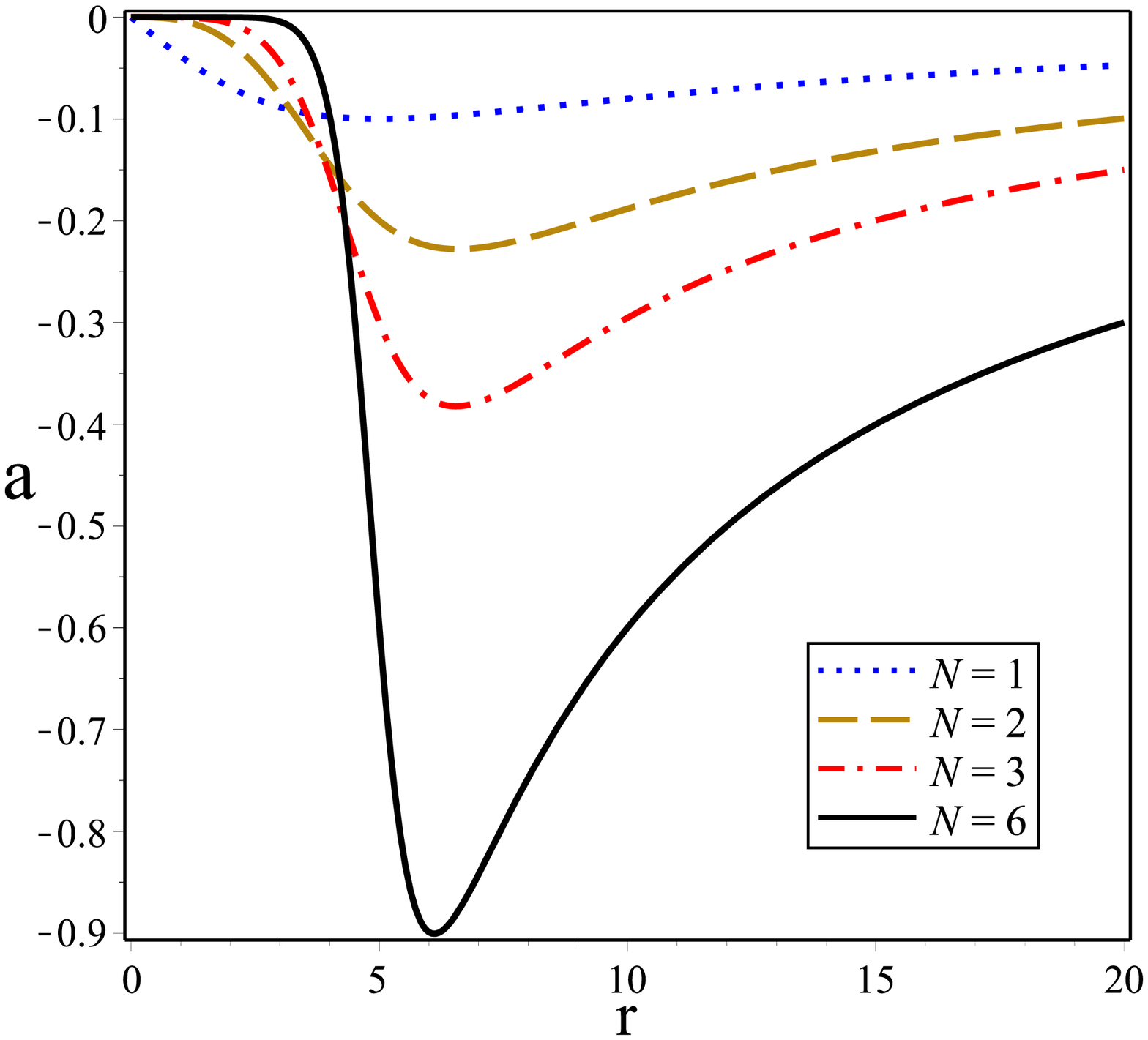}
\caption{Gauge field $a\left( r\right) $.}
\label{aa}
\end{figure}
\begin{figure}[tbp]
\centering\includegraphics[width=8.5cm]{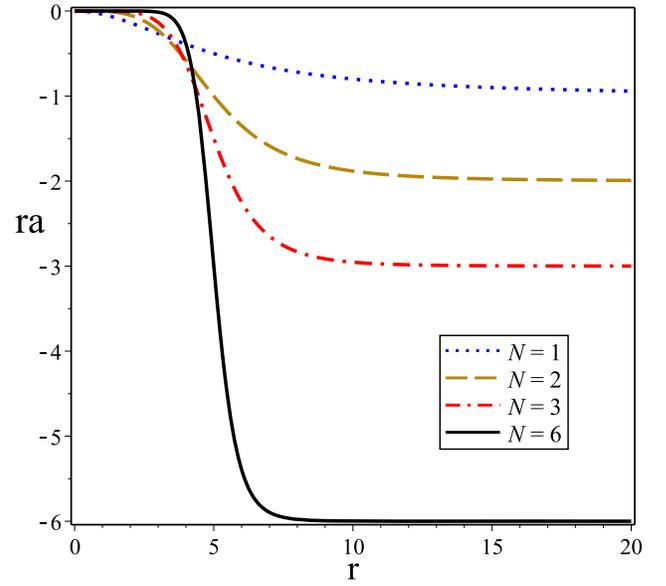}
\caption{$ra(r)$ profiles.}
\label{asas}
\end{figure}

The Fig. \ref{aa} depicts the profiles of the gauge field, the minimum is
located at $r=r_{0}\left( 2N-1\right) ^{1/2N}$ such that when $N$
increases its position close to the value $r_{0}$. The
Fig. \ref{asas} shows its asymptotic behavior as explicitly given in Eq. (%
\ref{r-inf}).

\begin{figure}[tbp]
\centering\includegraphics[width=8.5cm]{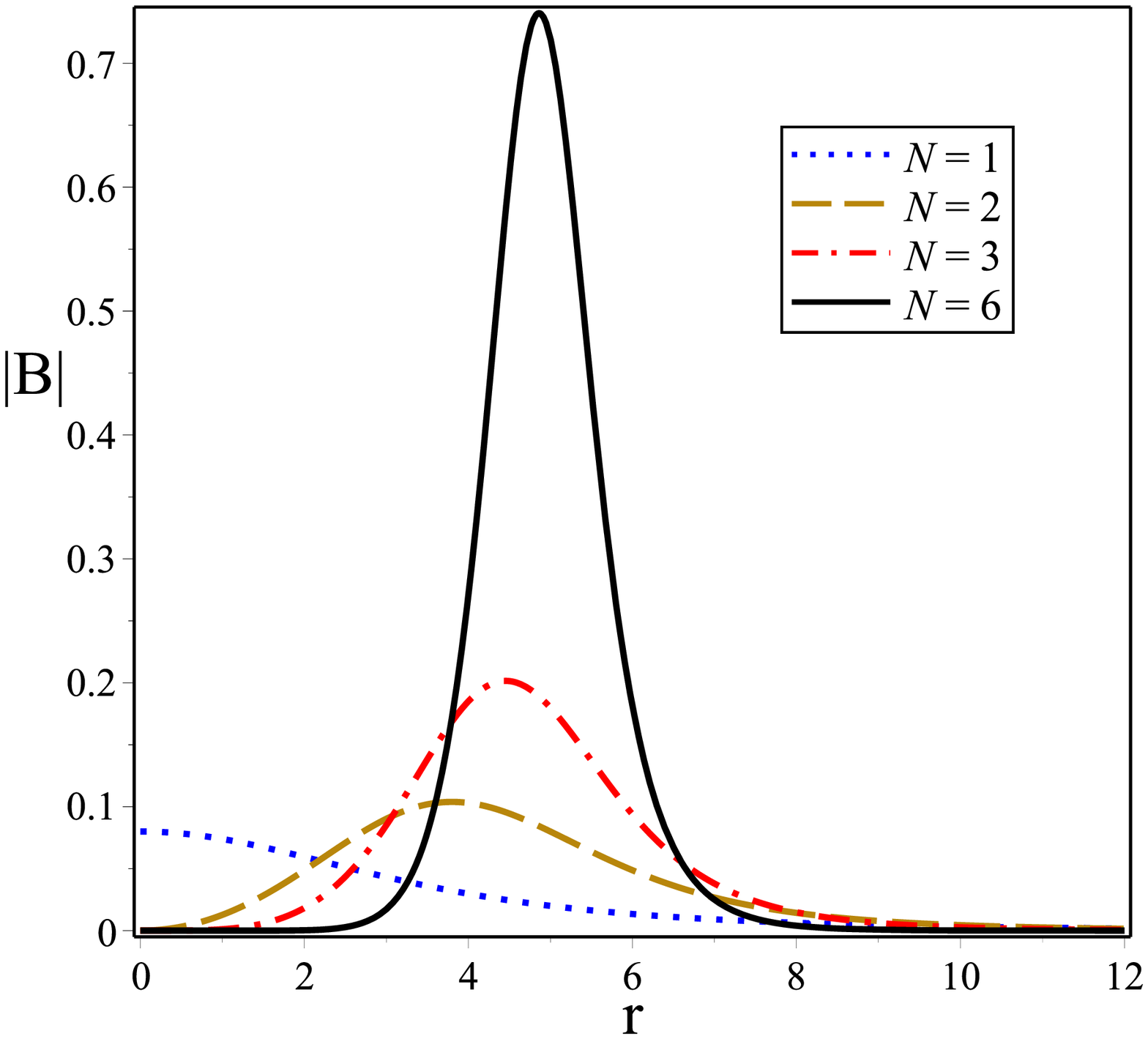}
\caption{Magnetic field $|B\left( r\right) |$.}
\label{mm}
\end{figure}

Fig. \ref{mm} depicts the profiles of the magnetic field. For $N=1$ it is a lump centered at origin but for $N>1$ its
maximum is located in $r=r_{0}\left( \displaystyle{\frac{N-1}{N+1}}\right)
^{1/2N}$. For large values of $N$, it is locate very close
to $r_{0}$ and its amplitude goes as $\displaystyle\frac{N^{2}}{%
2r_{0}^{2}}$.

\begin{figure}[tbp]
\centering\includegraphics[width=8.5cm]{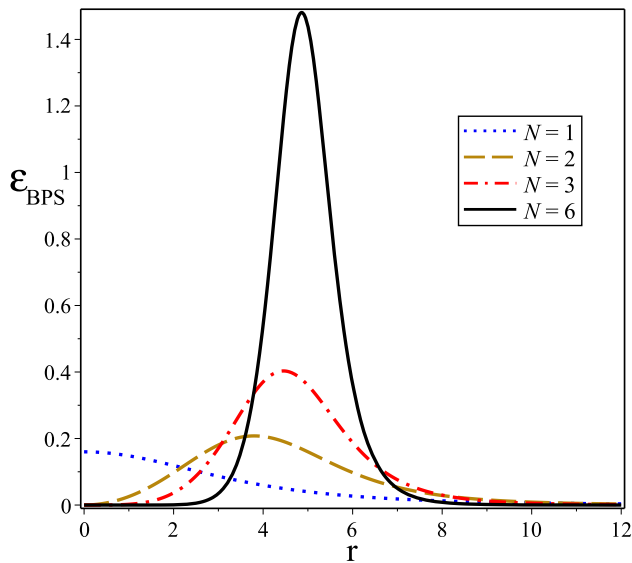}
\caption{BPS energy density $\protect\varepsilon _{BPS}\left( r\right) $.}
\label{xx}
\end{figure}

Fig. \ref{xx} shows the profiles of the BPS energy density which
have a similar behavior as the magnetic field.

\section{Remarks and conclusions}

We have analyzed a Chern-Simons CP(1) model with generalized kinetic
term. Such generalization allows to obtain self-duality equations whose
analytical solutions\ minimize the energy density. We have obtained a lower
bound for the BPS energy given by a sum of two contributions, the first one
is due to the magnetic flux and the second one is related to CP(1)
topological charge characterizing the BPS solutions. Because our self--dual
solutions provide quantized magnetic flux proportional to $N$,
the CP(1) topological charge (see Eqs. (\ref{fluxx}) and (\ref{ccn})), the
BPS\ energy result proportional to the CP(1)\ topological charge.

\subsection*{Acknowledgements}
We would like to thank the referee for a careful reading of the paper, as well as for the interesting remarks and comments.
RC thanks to CAPES, CNPq and FAPEMA (Brazilian agencies) for partial
financial support and LS thank to CAPES for full support.

\end{document}